\documentclass{ieeeaccess}

\usepackage{cite}
\usepackage{amsmath,amssymb,amsfonts}
\usepackage{algorithmic}
\usepackage{graphicx}
\usepackage{textcomp}
\usepackage{multirow}
\usepackage{caption}
\usepackage{tabularx}
\usepackage{array}

\begin{document}
\history{Date of publication xxxx 00, 0000, date of current version xxxx 00, 0000.}
\doi{10.1109/ACCESS.2017.DOI}

\title{Author Impact: Evaluations, Predictions, and Challenges}
\author{\uppercase{Fuli Zhang}\authorrefmark{1},
\uppercase{Xiaomei Bai}\authorrefmark{2},
\uppercase{Ivan Lee}\authorrefmark{3}}
\address[1]{Library, Anshan Normal University, Anshan 114007, China}
\address[2]{Computing Center, Anshan Normal University, Anshan 114007, China}
\address[3]{School of Information Technology and Mathematical Sciences, University of South Australia, Adelaide SA 5001, Australia}
\tfootnote{This work was partially supported by Liaoning Provincial Key R\&D Guidance Project (2018104021) and Liaoning Provincial Natural Fund Guidance Plan (20180550011).}


\corresp{Corresponding author: Xiaomei Bai (e-mail: xiaomeibai@outlook.com).}

\begin{abstract}
  Author impact evaluation and prediction play a key role in determining rewards, funding, and promotion. In this paper, we first introduce the background of author impact evaluation and prediction. Then, we review recent developments of author impact evaluation, including data collection, data pre-processing, data analysis, feature selection, algorithm design, and algorithm evaluation. Thirdly, we provide an in-depth literature review on author impact predictive models and common evaluation metrics. Finally, we look into the representative research issues, including author impact inflation, unified evaluation standards, academic success gene, identification of the origins of hot streaks, and higher-order academic networks analysis. This paper should help the researchers obtain a broader understanding in author impact evaluation and prediction, and provides future research directions.
\end{abstract}


\titlepgskip=-15pt

\maketitle


\section{Introduction}
\label{sec:introduction}
\PARstart{B}{ig} scholarly data has grown exponentially which aligns to the expansion of academic activities and productivity, however, it has also brought unprecedented challenges~\cite{xia2017big}. For example, it is difficult to identify most relevant research work or scholars from a vast amount of scholarly data through a simple search. In addition, for decision makers who allocate research funds, more information is needed to support the research evaluation system not only reflects past performance, but also predicts potential research productivity~\cite{bai2019predicting,cai2019scholarly,bai2016identifying}. Therefore, author impact evaluation and prediction are of great significance. On one hand, it is possible to distinguish authors' impact and provide assistance, especially for beginners, to explore a new research field. On the other hand, author impact evaluation provides support for rewards, funding, and promotion decisions to a certain extent.

The past few decades have witnessed the progress of research in author impact evaluation and prediction, including changes in research focuses: (1) from past performance analysis to future prediction of author impact; (2) from simple citation analysis to complex citation analysis; (3) from unstructured metrics to structured metrics; (4) from a single dimension of evaluation methods to multiple dimensions of evaluation methods.
To quantify scholarly impact, citation has been the most widely used technique~\cite{Eugene2006The,Pan2014Author,Farooq2017DS}. A large number of citation-based indicators are proposed, such as h-index and its variants~\cite{hirsch2005index,Egghe2006Theory,Alonso2009h,zhang2009index}. However, the method of measuring author impact from a single dimension has been unable to meet the rapid development of big scholarly data~\cite{Hirsch2007Does}. The emergence of academic media platforms and the evolution of social network relationships have challenged the evaluation and prediction of author impact~\cite{bai2018predicting}. Structured evaluation based on citations is a popular method for quantifying author impact in recent years~\cite{senanayake2015pagerank,Fiala2015Do,Pradhan2017}. This method evaluates author impact mainly from the perspective of scholarly network structure. The advantage of the network-based structured evaluation method is that it can use rich scholarly data and relationships in academic community rather than relying solely on citation relationships.

As an alternative to structured evaluation, model-based methods have also been introduced for author impact prediction~\cite{Sinatra2016Quantifying,Nezhadbiglari2016Early}. Sinatra et al.~\cite{Sinatra2016Quantifying} introduced a stochastic model which assigns a unique parameter $Q$ for each individual author, to accurately predict the evolution of the author's impact. The $Q$ model mainly considers the effects of productivity, individual ability, and luck, to form a generalised pattern of scientific success.

Although researchers have delivered various achievement in author impact evaluation and prediction, many challenging problems remains unresolved~\cite{Dunaiski2018Author,fiala2017pagerank,dunaiski2018evaluate,higham2017unraveling,panagopoulos2017detecting}. The heterogeneous attribute and the dynamic nature of big scholarly data lead to highly diversified scholarly networks, which raises the challenge in exploring the relationship between authors and other scholarly entities. At present, in most of the author impact prediction models, implicit features and implicit relationship mining need to be further improved, namely, factors that can influence the success of scholars need to be explored in depth. By achieving these, it will be possible to more accurately discover the academic rising star and more reasonably evaluate and predict author impact. Deep citation behavior analysis is another challenging issue in the existing relevant structured author impact evaluation and prediction research. The author's citation behavior is complex and diverse, and it is necessary to fully explore the hidden relationships in scholarly networks and fine-tune the evaluation and prediction models.

This paper presents a review of recent developments in author impact evaluation and prediction, and the review complements relevant work in the past: Waltman et al.~\cite{Waltman2016A} offer a review of the literature on citation impact indicators. This overview covers data sets, basic citation impact indicators, the topics of normalization, counting methods, journal citation impact indicators, and recommendations for future research. This overview has a broader scope than our presented overview, but it covers the most basic indicators such as citations, the number of highly cited publications, and h-index. Wildgaard et al.~\cite{Wildgaard2014A} present a review on author impact evaluation. One limitation of this review is that it does not consider author impact prediction research. In this paper, the progress of author impact evaluation and prediction is described in detail.

The rest of this paper is organized as follows. First, author impact evaluation method is presented in Section 2. Next, a review of the literature on author impact prediction method is provided in Section 3. Open issues and challenges are then discussed in Section 4. Finally, we conclude this paper in Section 5.

\section{Author impact evaluation}
\begin{figure*}[!htbp]
  \centering
  \includegraphics[width=0.88\linewidth]{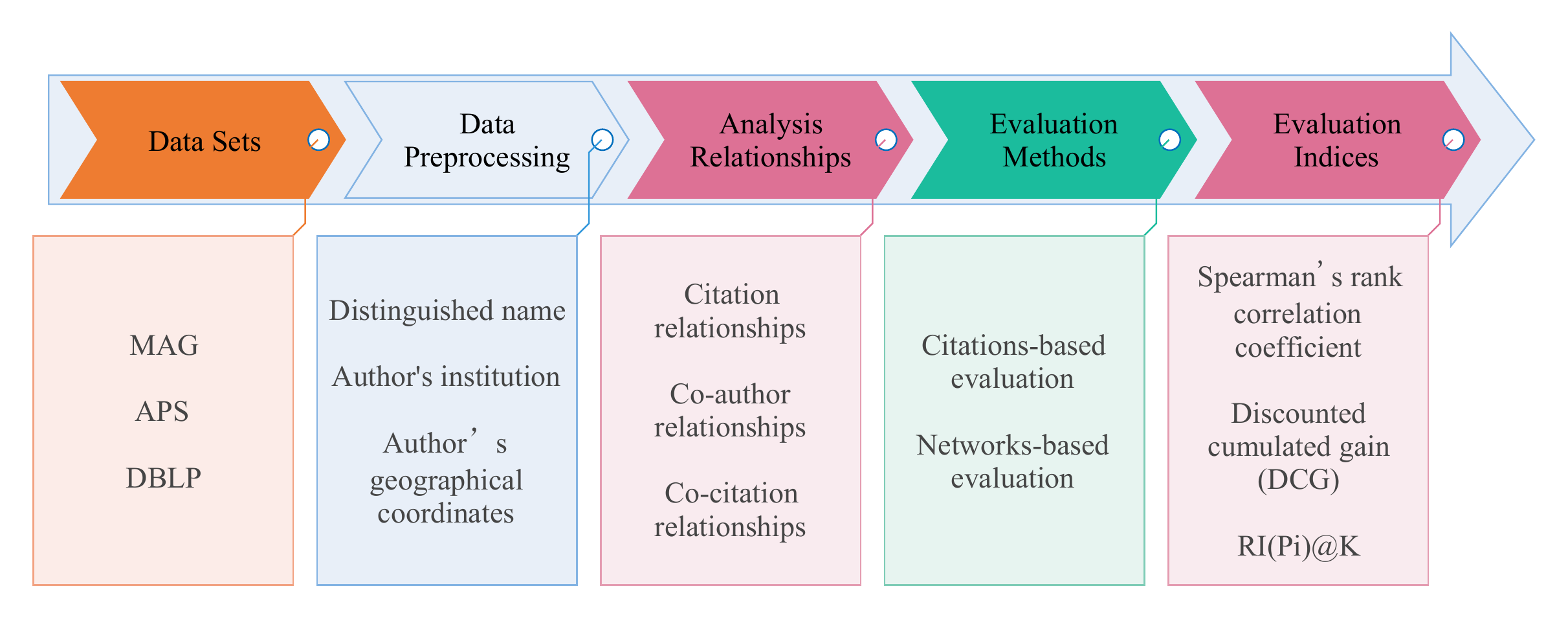}
  \caption{Framework of author impact evaluation.}
  \label{figure1}
\end{figure*}

Author impact research mainly addresses two related issues: (1) evaluate the past impact of authors; and (2) predict their future impact. Author impact evaluation includes the following parts: data collection, data pre-processing, data analysis, feature selection, algorithm design, and algorithm evaluation, as shown in Figure~\ref{figure1}.

\subsection{Data sources}
Web of Science, Scopus and Google Scholar are frequently used for author impact evaluation. Web of Science and Scopus are subscription-based databases. In addition to covering journals and book series, Web of Science also offers conference proceedings citation index~\cite{Waltman2016A}. More scholarly resources can be retrieved by Google Scholar, including meta data of scholarly papers, conference proceeding, books, theses, patents and technical reports.

In addition to proprietary data sets, several public accessible data sets are available, including American Physical Society (APS)\footnote{http://publish.aps.org}, Digital Bibliography \& Library Project (DBLP)\footnote{https://dblp.uni-trier.de/}, and Microsoft Academic Graph (MAG)\footnote{http://aka.ms/academicgraph}. One advantage of APS is that it provides citation records as part of its data set. DBLP has distinguished different authors based on names, but it does not provide citation records. In comparison, MAG offers heterogeneous information with publication records, authors, institutions, journals, conferences, fields of study and citation relationships.

Apart from accessing meta records made available in proprietary or public-accessible data sets, another approach is to crawl some online social data such as downloads, mentions, tweets, shares, views, discussions, saves, and bookmarks for author impact evaluation~\cite{xia2016bibliographic}.

\subsection{Data pre-processing}
Data pre-processing is crucial for author impact evaluation as it significantly impacts the accuracy. Upon obtaining the author's raw data, the few questions need to be considered: (1) How to accurately differentiate authors based on names and affiliation? (2) How to account for authors who are associated to multiple affiliations? (3) How to weight individual author contributions in jointly published papers?

In practice, different pre-processing techniques are taken place subject to different evaluation objectives. For example, in author impact evaluation and prediction research, author name disambiguation is necessary to distinguish authors with same full names for some datasets~\cite{ferreira2012brief}, such as the APS dataset which is commonly used for scholarly data analysis in the Physics discipline.

\subsection{Data analysis}
In author impact evaluation, scholarly data analysis can be divided into two categories: statistical analysis and scholarly network analysis. Statistical analysis can reveal the scientific knowledge behind the big scholarly data by using statistical analysis~\cite{xia2017big}.

The heterogeneity and diversity attribute of scholarly network structure have raised the challenges in scholarly network analysis. In recent decades, researchers have made important progress in network analysis research, such as the structural hole theory~\cite{ahuja2000collaboration}. The structural hole theory has been applied to academic networks by researchers to evaluate author impact~\cite{lou2013mining,zhang2018airank}. Their experimental results indicate the structural hole has a very close relationship to individual scholar's success. Social network connecting authors and co-authorship network have attracted increasing attention for author impact prediction. Zhou et al.~\cite{zhou2007co} propose a co-ranking method to evaluate authors and their publications based on three scholarly networks: authors' social network, citation network, and co-authorship network.

\subsection{Feature selection}
Early studies in author impact evaluation mainly consider two quantitative features: citation counts and paper counts. Recent researches have been inspired from the PageRank algorithm and have used the structural features of scholarly networks to assess the author impact~\cite{dunaiski2016evaluating,falagas2008comparison,ding2009pagerank,yan2011discovering}. Also, social network measures such as degree centrality, closeness centrality, betweenness centrality, and PageRank frequently are used to assess author impact~\cite{bollen2009principal,ding2009pagerank,zhang2017exploring,Dunaiski2018Author,Dalibor2012Time,nykl2015author,usmani2017unified,dunaiski2016evaluating,senanayake2015pagerank}. In addition, Tweets are used to quantify author impact~\cite{kong2011tweet}. Table~\ref{tab:1} shows an example of selected feature for evaluating author impact.
\begin{table*}[!htbp]
  \centering
  \caption{An example of selected features for evaluating author impact.}
  \begin{tabular}{|p{8cm}<{\centering}|p{3cm}<{\centering}|p{3cm}<{\centering}|}
    \hline
    Features&Feature Types&References\\ \hline
    citations &statistical feature& \cite{ding2009pagerank,bharathi2013evaluation,dorogovtsev2015ranking,Jiang2013Graph,usmani2017unified,marchant2009score,Sinatra2016Quantifying,Bouyssou2016Ranking}\\ \hline
    maximum entropy &statistical feature& \cite{stallings2013determining}\\ \hline
    time &statistical feature& \cite{zhang2017exploring}\\ \hline
    the number of papers published, the number of co-authors &statistical feature& \cite{usmani2017unified,stallings2013determining}\\ \hline
    the number of authors, the number of papers in a certain journal &statistical feature& \cite{marchant2009score}\\ \hline
    the active year, the citing times, the citing times per paper, and the citing times per year, the cited time of papers one author cites, the citing time of papers one author cites, and the cited time of co-authors &statistical feature& \cite{stallings2013determining}\\ \hline
    references being cited by the author before and its ratio, references in the author's previous publications and its ratio, keywords, the times of author attend venue &statistical feature& \cite{zhang2018task}\\ \hline
    number of unique publication venue, number of paper-paper citation edges, number of coauthorship edges, number of author-citation edges&statistical feature& \cite{Pradhan2017}\\ \hline
    tweets &statistical feature& \cite{kong2011tweet}\\ \hline
    bridge counts, betweenness, diversity of cooperators&statistical feature & \cite{zhang2018airank}\\ \hline
    degree, betweenness, closeness&network features & \cite{ding2009pagerank}\\ \hline
    PageRank&network features & \cite{ding2009pagerank,zhang2017exploring,Dunaiski2018Author,Dalibor2012Time,nykl2015author,usmani2017unified,dunaiski2016evaluating,senanayake2015pagerank}\\ \hline

    network index(NI), PageRank, HITS&network features& \cite{zhang2018airank}\\ \hline
    researcher importance&network features& \cite{Jiang2013Graph}\\ \hline
    Eigenfactor scores&network features& \cite{west2013author}\\ \hline
    paper authority vector&network features& \cite{liang2016scientific}\\ \hline
  \end{tabular}
  \label{tab:1}
\end{table*}

\subsection{Author impact evaluation}

\subsubsection{Citations-based evaluation}
The most representative of author impact evaluation method is h-index, which defines that ``a scientist has index $h$ if $h$ of his or her $N_{p}$ papers have at least $h$ citations each and the other ($N_{p}-h$) papers have $\leq h$ citations each''~\cite{hirsch2005index}. While the simplicity of h-index might be the reason for its popularity~\cite{franceschini2010analysis}, some researchers have pointed out its drawbacks. For instance, h-index is a cumulative measure so it does not fall, thus it fails to reflect the reduced impact for scholars that become inactive in research. Also, h-index does not differentiate different citations such as self-citations by default, thus the measure may not reflect true impact subject to different degrees of manipulation~\cite{engqvist2008h}.

With the limitations in h-index, a large number of h-index variants have emerged to address these shortcomings~\cite{Bornmann2014Are,egghe2008h,egghe2006improvement,bornmann2009there,guns2009real,bornmann2009we,zhang2009index}.
Table~\ref{tab:2} compares h-index and its representative variants. Egghe~\cite{Egghe2006Theory} proposes g-index, which not only keeps all advantages of the h-index, but also measures the global citation performance of an author. If the number of citations of an author's published papers is in descending order, the g-index is the largest number that top $g$ papers received at least $g^{2}$ citations. Since g-index better takes into account the number of citations of top papers of an author, g-index is easily affected by highly cited papers. For example, an author publishes $11$ papers, in these papers, the number of citations of just one paper is $100$, the number of citations of other $10$ papers is equal to $1$. Although the author' h-index is 1, the author's g-index is $10$. To overcome the shortcomings of g-index, Alonso et al.~\cite{Alonso2010hg} quantify each author impact by using their proposed hg-index (see hg-index formula in Table~\ref{tab:2}), which keeps the advantages of h-index and g-index and minimizes their disadvantages. For the above example, the author's hg-index is $\sqrt{10}$. By comparing hg-index, g-index, and h-index, the advantage of hg-index is obvious and is listed as follows: (1) hg-index weakens the impact of highly cited papers; (2) hg-index solves the shortcoming of h-index. For the above example, if the author impact uses the w-index~\cite{wu2010w} to assess, the author's w-index is also $1$, it equals to the value of h-index. In this way, w-index does not solve the shortcoming of h-index to a certain extent. Due to the ignored excess citations of h-index, Zhang~\cite{zhang2009index} defines e-index. The excess citations received by all publications in the h-core can be denoted by $e^{2}$. The e-index is a necessary complement for h-index, especially for assessing highly cited scholars. Further, to overcome the limitation of h-index and e-index, Bihari et al.~\cite{bihari2017index} propose EM-index, which is the extension of h-index and e-index. EM-index is more fine-grained indicator than the h-index, g-index, and e-index. However, EM-index does not consider the citations of  all publications. Therefore, the EM-index is suitable to evaluate the highly cited papers. To overcome this limitation of EM-index, a multidimensional extension of the EM-index called EM'-index is proposed~\cite{bihari2017index}.  Subsequently, to overcome this limitation of year based indices, Bihari~\cite{bihari2018year} defines the year based EM-index and the year based EM'-index. In their methods, they consider three different parameters: the total number of papers, the year citations of paper, and citations obtained in a particular year.


In addition, Eggle et al.~\cite{egghe2008h} propose a weighted h-index, named $h_{w}$-index, depending on the obtained citations of papers belonging to the h-core, and $h$-index $\leq$ $h_{w}$-index $<$ $g$-index. W{\"u}rtz et al.~\cite{wurtz2017stratified} propose the stratified h-index, which supplements the conventional h-index in three separate h-indices: first authorships, second authorships, and last authorships. Other indices, such as Multiple h-index~\cite{yaminfirooz2015multiple}, rp-index and cp-index~\cite{Abbasi2010Evaluating}, b-index~\cite{brown2009simple}, $q^{2}$-index~\cite{cabrerizo2010q2}, year-based h-type indicators~\cite{mahbuba2013year}, pure h-index~\cite{wan2007pure}, Wl-index~\cite{wan2014wl}, R-and AR-indices ~\cite{jin2007r}, $\pi$-index~\cite{vinkler2009pi}, and $h^{m}$-index~\cite{schreiber2008modification} are used for author impact evaluation.

\begin{table*}[!htbp]
  \centering
  \caption{Comparing h-index and its representative variants.}
  \begin{tabular}{|p{1.8cm}<{\centering}|p{4cm}<{\centering}|p{4cm}<{\centering}|p{4cm}<{\centering}|}
    \hline
    Method and reference&Formula&Advantage&Disadvantage\\ \hline
    h-index~\cite{hirsch2005index}& 
     $h=\frac{c}{1+\frac{c}{p}}\cdot n$, where $c$ is new citations per year every subsequent year, $p$ is papers of researcher published per year, $n$ is $n$ year& an easily computable index, it gives an estimate of the importance, significance for an author's publications & h-index is influenced by self-citations, h-index will not decrease, excess citations are completely ignored, does not quantify co-authors' contributes \\ \hline
    g-index~\cite{Egghe2006Theory}& $g=\left (\frac{\alpha -1}{\alpha -2}  \right )^{\frac{\alpha -1}{\alpha}}\cdot h$, where $\alpha$ is a parameter, $h$ is the value of h-index &  as simple as the h-index, better takes into account the citation scores of the top articles& the g-index may be greatly influenced by a very successful paper \\ \hline
    hg-index~\cite{Alonso2010hg}&$hg=\sqrt{h\cdot g}$, where $h$ is h-index, $g$ is g-index & simple to compute, provides more granularity than the h- and g- indices, takes into account the cites of the highly cited papers, significantly reduces the impact of single very high cited papers&  hg-index is influenced by self-citations \\ \hline
    w-index~\cite{wu2010w}& $ w=max(cit_{p}\geq w-p+1), for \quad all \quad p\leq w$, where $cit_{p}$ is the citations of the $p$-th paper &  as simple as the h-index, the w-index is the largest isosceles right angle triangle under the citation curve&  w-index is influenced by self-citations \\ \hline
    $h_{w}$-index~\cite{egghe2008h}& $ w=\sqrt{\sum_{j=1}^{k}cit_{j}}$, where $cit$=citation counts, $j$ is the largest row index &  depending on the obtained citations of papers belonging to the h-core&  this index is influenced by self-citations, it does not consider the excess citation count\\ \hline
    e-index~\cite{zhang2009index}& $ e=\sqrt{\sum_{p=1}^{h}\left ( cit_{p}-h \right )}$, where $h$ is the h-index, $cit_{p}$ is the citation count of pth paper
                                   &  excess citations are considered, especially for evaluating highly cited scientists&  it does not consider the core citation count \\ \hline
    $EM$-index~\cite{bihari2017index}& $EM=\sqrt{\sum_{e=1}^{k}E_{e}}$, where $EM$ is the EM-index of an author, $E_{1}$ =h, $E_{2}$ can be obtained by calculating the h-index from the excess citations of h-core paper  & considers the core citation count and the excess citation count&this index does not consider the citations of all publication \\ \hline
    $EM'$-index\quad ~\cite{bihari2017index}& $EM'=\sqrt{\sum_{e=1}^{k}E_{e}'}$, where $E'$ is the k-dimensional 
    vector that contains the citations of all the papers (these papers are cited at least once)
    & The $EM'$-index is the multidimensional extension of the $EM$-index, consider all papers& The $EM'$-index is influenced by self-citations \\ \hline
    Year based $EM$-index~\cite{bihari2018year}& $Y\_EM=\sqrt{\sum_{i=1}^{k}YE_{e}}$,
    the year based EM-index is the square root of the sum of the component of year based EM-index & are the extension of year based h-indices &this index does not consider the citations of all publication \\ \hline
    Year based $EM'$-index\quad~\cite{bihari2018year}& $Y\_EM'=\sqrt{\sum_{e=1}^{k}YE_{e}'}$,  
    the year based  $EM'$-index is the square root of the sum of the component of year based  $EM'$-index & year based $EM'$-index considers all the items which occur at least one time& this index is influenced by self-citations \\ \hline
  \end{tabular}
  \label{tab:2}
\end{table*}

\subsubsection{Network-based evaluation}
Because citations may be easily manipulated, citation-based indices may not objectively evaluate the actual impact of authors. Instead of citations, scholarly networks are used for author impact evaluation. Network-based methods have been investigated as alternative methods for author impact evaluation.

\begin{figure*}[!htbp]
  \centering
  \includegraphics[width=0.88\linewidth]{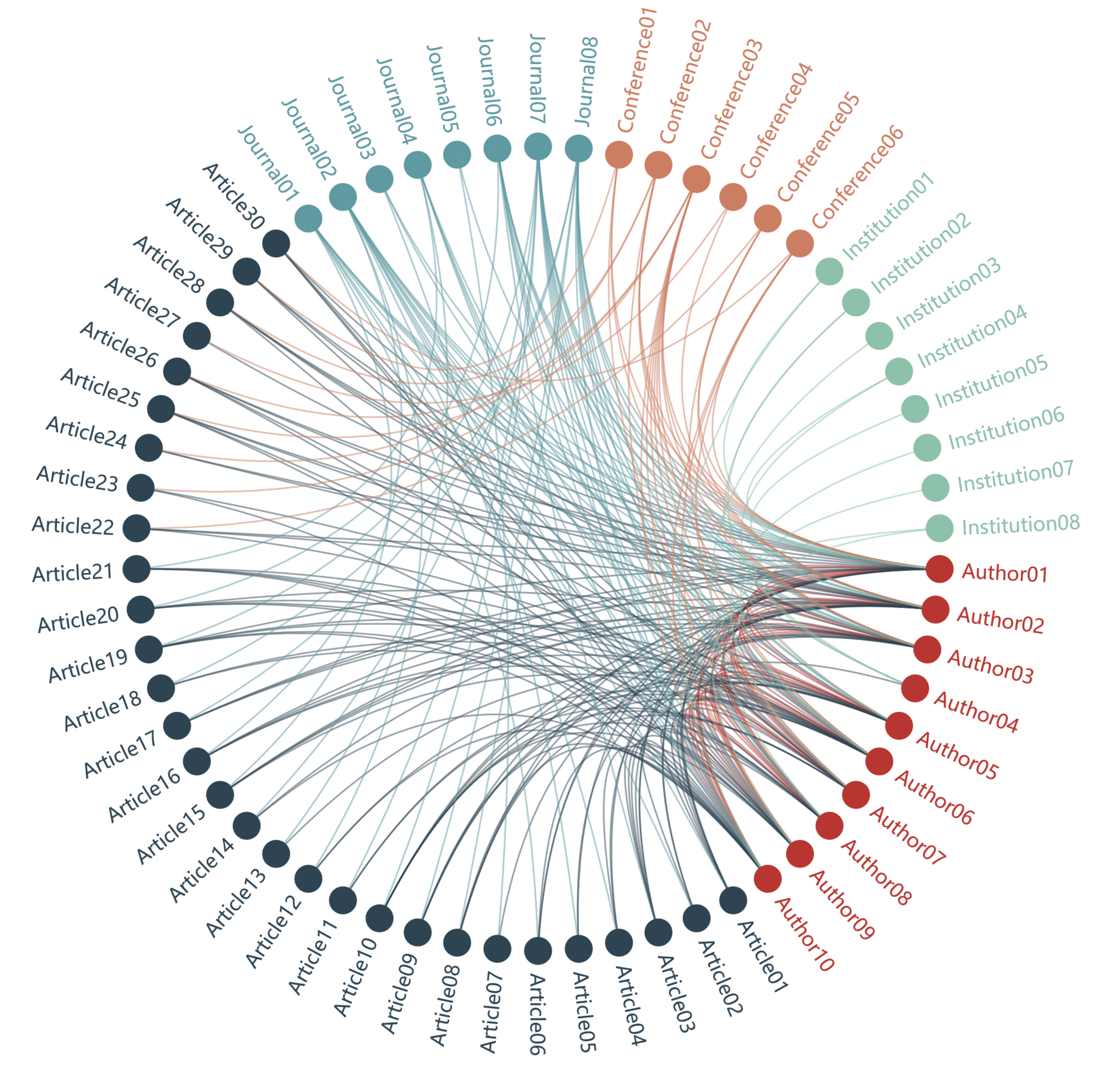}
  \caption{Eight typical scholarly networks - an example of 10 randomly-selected computer science authors from the MAG data set.}
  \label{figure2}
\end{figure*}

The exponential growth of academic data offers unprecedented opportunities to explore patterns characterizing the structure of scholarly networks and evolution of science~\cite{fortunato2018science}. To demonstrate these academic relationships (see Figure~\ref{figure2}), we randomly selected 10 authors from computer science area in the MAG dataset to construct eight typical networks based on papers they published, journals or conferences, and institutions. These scholarly networks include citation network, co-author network, author-paper network, author-journal network, author-institution network, author-conference network, paper-journal network, and paper-conference network. In Figure~\ref{figure2}, different color nodes represent different types of academic entities and the lines between them represent scholarly relationships. Because the shortcomings exist in author impact evaluation based on citations, as mentioned in Table~\ref{tab:2}, researchers measure author impact by using scholarly networks. By exploring quantitative methods, from statistics to network science approaches, machine learning algorithms and mathematical analysis, scientists have developed structural author impact evaluation methods based on scholarly networks (see Table~\ref{tab:3}).

Table~\ref{tab:3} compares the different author impact evaluation methods in the following eight aspects, including method and reference, scholarly network, homogeneous relationships, heterogeneous relationships, data sets, comparing algorithms, evaluation metrics, and performance. Ding et al.~\cite{ding2009pagerank} introduce PageRank algorithm to academic network, and its purpose is to use the PageRank algorithm to evaluate the author impact. During this period, researchers mainly leverage homogeneous networks for evaluation. Based on PageRank, Pradhan et al.~\cite{pradhan2016c} propose $C^{3}$-index, which ranks authors by using the weighted multi-layered scholarly networks, including author-author citation network, author-author co-authorship network, and paper-paper citation network. The $C^{3}$-index score can be obtain by computing three individual component scores from three layers mentioned above. The component scores for individual layer are actually the PageRank scores.

Recently, author impact evaluation has received wide attention, especially heterogenous network with multiple types of nodes and relationships. Liu et al.~\cite{liu2014tri} propose a graph-based ranking framework, Tri-Rank, to co-rank authors, scholarly papers, and venues simultaneously in heterogenous scholarly networks. Their experimental results show that Tri-Rank with heterogenous networks is more effective and efficient than PageRank~\cite{page1998pagerank}, HITS~\cite{kleinberg1999web}, and Co-Rank~\cite{zhou2007co} in ranking authors. However, in these researchers, all citations are regarded as equal importance. To automatically identify how references in a bibliography affect on the citing paper, Zhu et al.~\cite{zhu2015measuring} examine the effectiveness of several features to determine the academic influence of a citation. Furthermore, researchers consider weighted citation networks to measure scholar impact. Nykl et al.~\cite{nykl2015author} use h-index, the number of papers, citations, journal impact values and author count of scholarly paper features as citation weight in citation networks. Further, they apply PageRank algorithm in the citation networks. Their experimental results indicate that using the journal impact values in PageRank can improve author ranking. Li et al.~\cite{li2014network} propose a network-based and multi-parameter model to find influential authors. The idea stems from the fact that the authority of scholarly networks changes as nodes are removed. Author $i$'s prestige in academic networks is defined as
\begin{equation}\label{eq:e1}
  p_i(g) = \alpha_{i} \cdot \sum_{j=1}^{n}b_{ij}(g,\beta)
\end{equation}
where $b_{ij}(g,\beta)$ represents the element of matrix $B(g,\beta)$ at row $i$ and column $j$. The parameter $\alpha$ represents the base value of node. The parameter $\beta$ can capture the value being connected to certain node, which decays with distance.

Citation networks evolve over time, thus time represents an important feature to quantify scholarly or institutional impact~\cite{bai2017implicit}. Wang et al.~\cite{wang2013ranking} define a time-aware weights of edges strategy for evaluating scholarly impact. In practice, they find that older publications can get more accurate predictions than recent ones. Therefore, they give the edges associated with older authority publications higher weights, because the scores of these publications are more reliable than those of new publications. In their researches, the hub scores of an author can be obtained by
\begin{equation}\label{eq:e2}
  H(A_i) = \frac{\sum_{P_{j}\in Neighbor( A_{i} )}W_{ap}(i,j)\cdot S( P_{j} )}{ \sum_{P_{j}\in Neighbor(A_{i})}W_{ap}(i,j)}
\end{equation}
where $Neighbor(A_{i})$ is the collection of papers, neighboring to author $A_{i}$. $S( P_{j})$ is the score of scholarly paper, and $W_{ap}(i,j)$ is the weight of edge from author $A_{i}$ to paper $P_{j}$. The weight $W_{ap}(i,j)$ can be calculated by
\begin{equation}\label{eq:e3}
  W_{ap}(i,j) = a^{T_{current}-T_{i}}
\end{equation}
where $T_{current}-T_{i}$ indicates the age (in years) of the paper $P_{i}$ since it was published. $a$ is a constant with its value greater than 1. In their experiments, the value of $a$ is set as 2. In addition, a temporal citation network among scholars is used by Franceschet et al.~\cite{franceschet2017timerank}, who allocate ratings by considering the relative position between two authors at the time of the citations. Thus, they name the dynamic rating method as TimeRank. The difference between TimeRank and PageRank is that PageRank uses the citing author's absolute rating, while TimeRank uses the citing author's relative rating. It is worth mentioning that the ratings of citing author for PageRank are at the end of the temporal citation, while TimeRank uses the ratings of citing and cited scholars at the actual time of citation.

Apart from the time factor, scholars' position in the academic networks may also be utilized to assess the scholar impact. Zhang et al.~\cite{zhang2018airank} first define the scholar's structural index (SI) to capture the effect of scholars' positions in scholarly networks. They then use the PageRank and HITS algorithms together to obtain the scholar's network index (NI). Finally, based on the values of SI and NI, they calculate scholar's final score. In their research, to determine scholars' positions in scholarly network, they apply the structural holes theory which indicates that scholars linking different disciplines have more influence.

The evaluation models described above suffer from one limitation, namely they are usually based on scholarly network structure, ignoring the content semantic awareness. To address this limitation, Zhang et al.~\cite{zhang2018task} propose a task-guided and semantic-aware ranking model. The ranking model performs joint optimization of GRU-based content encoding and task-guided ranking. Their experimental results demonstrate that the performance of TSR+ is better than a number of baselines.

\begin{table*}[!htbp]
  \centering
  \caption{Comparison of different network-based author impact evaluation methods.}
  \begin{tabular}{|m{2cm}<{\centering}|m{1.5cm}<{\centering}|m{1.5cm}<{\centering}|m{1.5cm}<{\centering}|m{1.5cm}<{\centering}|m{1.5cm}<{\centering}|m{1.5cm}<{\centering}|m{3cm}<{\centering}|}
    \hline
    Method and reference& scholarly network &Homogeneous relationships &Heterogeneous relationships&Data sets&Comparing algorithms&Evaluation metrics&Performance\\ \hline
    PageRank for ranking author~\cite{ding2009pagerank}&author-author& yes& no &Web of Science& PageRank with different damping factors& Spearman correlation coefficient& citation rank is highly correlated with PageRank with different damping factors\\ \hline
    P-Rank~\cite{zhao2009p}&paper-paper, co-author, paper-journal& yes& yes & DBLP& SimRank& Similarity &  The advantages of P-Rank are its semantic completeness, robustness and flexibility\\ \hline
    graph-based algorithms~\cite{gollapalli2011ranking}&author-paper& no& yes &ArnetMiner and UvT &baseline method&MAP, MRR& currently focusing on improving the expert ranking performance \\ \hline
    p-index~\cite{senanayake2014p}&paper-paper& yes& no & Google Scholar& h-index& & p-index is robust against manipulations and performs fairer and more effectively in ranking scientists\\ \hline

    Tri-Rank~\cite{liu2014tri}&author-paper, paper-venue, venue-author & no& yes &ACM Digital Library &PageRank, HITS, Co-Rank&Precision@k, Bpref, DCG& more effective and accurate than the state-of-the-art competitors including PageRank, HITS and Co-Rank\\ \hline
    $C^{3}$-index~\cite{pradhan2016c}&author-author, paper-paper, co-author& yes& no &MAS &h-index& Spearman correlation coefficient& $C^{3}$-index is as efficient as h-index\\ \hline
    TimeRank~\cite{franceschet2017timerank}&author-paper&  no& yes & Web of Science&TotCit, PageRank, h-index&Frequency& more effective than alternatives \\ \hline
    TRank (TR-re)~\cite{zhang2017exploring}&author-paper&  no& yes & DBLP and APS&RW, PRW, TR-ex, TR-po, PAve& AUC&better than a number of baselines \\ \hline

    $AIRank^{BrC}$~\cite{zhang2018airank}&co-author&  no& yes &MAG& $AIRank^{BeC}$ , $SI^{BrC}$
                                                                    , $SI^{BeC}$,NI, PageRank&AUC&better than other methods in ranking top scholars with more cross-domain citations\\ \hline
    task-guided and semantic-aware ranking (TSR+) model~\cite{zhang2018task}&author-paper&  no& yes & AMiner&TSR, TaskE, metapath2vec, word2v+BPR&Precision@k, Recall@k and AUC&better than a number of baselines\\ \hline
  \end{tabular}
  \label{tab:3}
\end{table*}

\subsection{Evaluation metrics}
Two popular metrics: Precision and Recall are usually used to evaluate the performance of author impact methods. The Precision shows the accuracy top-k authors by a method and it is calculated by $Precision=\frac{TP}{TP+FP}$, where $TP$ (True Positive) represents the number of positive cases that are correctly divided. $FP$ (False Positive) represents the number of positive cases wrongly divided. The Recall reflects the ration of true correlated authors returned in the top-k list. It is defined as: $Recall=\frac{TP}{TP+FN}$, where $FN$ represents the number of negative cases wrongly divided. In addition, Spearman's rank correlation coefficient and Discounted Cumulative Gain can be used to evaluate author impact.\\
\textbf{\emph {Spearman's rank correlation coefficient}}.\\
\begin{equation}\label{eq:e4}
  \rho =\frac{\sum_{i}(R_{1}(A_{i})-\overline{R}_{1})(R_{2}(A_{i})-\overline{R}_{2})}
  {\sqrt{\sum_{i}(R_{1}(A_{i})-\overline{R}_{1})^{2}\sum_{i}(R_{2}(A_{i})-\overline{R}_{2})^{2}}}
\end{equation}
where $R_1(A_i)$ and $R_2(A_i)$ are the position of author $A_i$ in the ground truth rank list and the corresponding algorithm rank list, respectively. $\overline{R}_{1}$ and $\overline{R}_{2}$ are the average rank positions of all authors in the two ranks lists, respectively.\\
\textbf{\emph {Discounted Cumulative Gain (DCG)}}.\\
\begin{equation}\label{eq:e5}
  DCG_{n}=\sum_{i=1}^{n}\frac{rel_{i}}{log_{2}{i+1}},
\end{equation}
where $DCG_{n}$ is the weighted sum of relevant degree of ranked authors, and its weight is a decreasing function varying according to the ranked position. Variable $i$ is the ranking of an author, and $rel_{i}$ is the relevance score of the $i$-th ranked author.\\

\section{Author impact prediction}
\begin{figure*}[!htbp]
  \centering
  \includegraphics[width=0.88\linewidth]{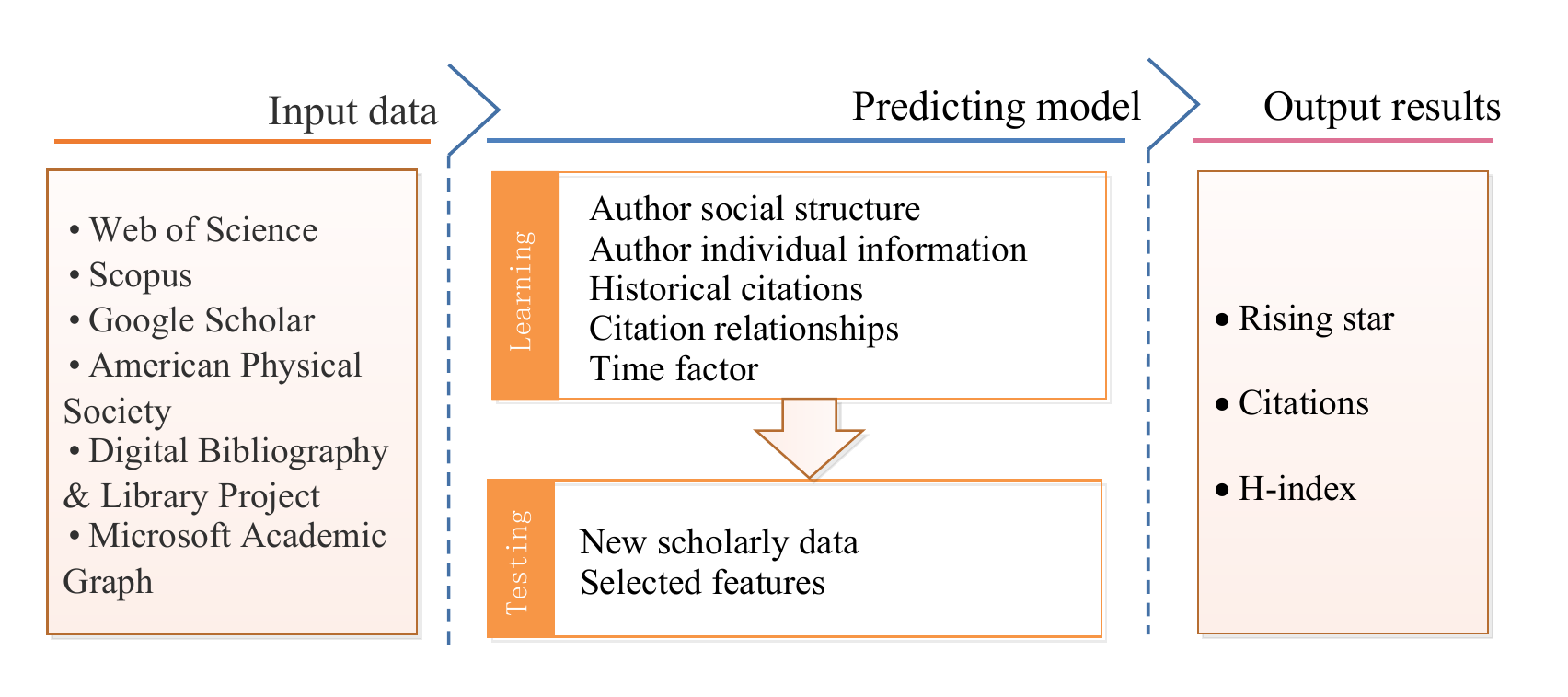}
  \caption{Framework of author impact prediction.}
  \label{figure3}
\end{figure*}

In the previous section, we have discussed author impact evaluation methods and common evaluation metrics. In this section, we focus specifically on author impact prediction models and common evaluation indices. The author impact predictive model can be roughly divided into three categories: feature-driven predictive model, network-based predictive model, and generative predictive model. The framework of author impact prediction includes input data, predictive model, and output results, as shown in Figure~\ref{figure3}.

\subsection{Author impact prediction model}
\subsubsection{Feature-driven predictive model}
To examine how one's h-index will evolve over time, feature-driven predictive models based on the following feature have been studied: author feature (author influence, number of co-authors, first author's h-index, average h-index of all authors), paper feature (citations, average citations, topic novelty, topic diversity), social feature (PageRank score, weighted average h-index of co-authors, co-author's citations) and other features (venue citation, venue count, venue specificity score)~\cite{mccarty2013predicting,dong2015will,daud2015using,dong2016can,daud2017finding,ayaz2018predicting}. Several representative examples about author impact prediction using mixture of features are summarized in Table \ref{tab:4}. In order to analyze the efficiency of multi-feature for author impact prediction, regression models are often used, such as linear regression~\cite{zou2003correlation}, semi-continuous regression~\cite{sohrabi2017effect} and XGboost~\cite{chen2016xgboost}.

\begin{table*}[!htbp]
  \centering
  \caption{Comparison of several representative multiple features-based author impact prediction models.}
  \begin{tabular}{|m{1cm}<{\centering}|m{1.8cm}<{\centering}|m{1.8cm}<{\centering}|m{1.8cm}<{\centering}|m{1.8cm}<{\centering}|m{1.8cm}<{\centering}|m{1.8cm}<{\centering}|m{1.5cm}<{\centering}|}
    \hline
    Reference& Author features &Paper features &Social features&Other features&Model&Evaluation metrics&Predictive target\\ \hline
    ~\cite{mccarty2013predicting}&number of co-authors, AvgAuthors, proportion of academic co-authors & &components, isolates, betweenness, hierarchy, MeanTie &&bivariate models, multivariate model, final model &$R^{2} coefficient $&h-index\\ \hline
    ~\cite{dong2015will}&A-first-max, A-ave-max, A-sum-max, A-first-ratio, A-max-ratio, A-num-authors, A-num-first&C-popularity, C-popularity-ratio, C-novelty, C-diversity, C-authority-first, C-authority-max, C-authority-ave  & S-degree, S-pagerank, S-h-co-author, S-h-weight  &R-ratio-max, R-citation, T-ave-h, T-max-h, T-h-first, T-h-max, V-ratio-max, V-citation & logistic regression classifier (LRC), random forest (RF), bagged decision trees (BAG)& precision, recall, F1-score, area under curve (AUC), accuracy &h-index\\ \hline
    ~\cite{daud2015using}&author influence, author contribution, temporal dimension& &co-author citations, co-author count&venue count, venue score, venue citation&MEMM, CART, BN, NB& average F1 score& rising stars\\ \hline
    ~\cite{dong2016can}&A-first-max,A-ave-max, A-sum-max, A-first-ratio, A-max-ratio, A-num-authors&C-popularity, C-novelty, C-diversity, C-authority-first, C-authority-max, C-authority-ave &S-degree, S-pagerank, S-h-co-author, S-h-weight& R-ratio-max, R-citation, T-ave-h, T-max-h, T-h-first, T-h-max,V-ratio-max, V-citation &LRC, SVM, NB, RBF, BAG, RF&precision, recall, F1-score, AUC, accuracy, MAP, Pre@3 &h-index\\ \hline

    ~\cite{weihs2017learning}& h-index, h-index variation over the last two years, cumulative citation count &  number of papers published, number of papers published in last two years & PageRank of author in unweighed co-authorship network, PageRank of author in weighed co-authorship network, & h-index of venues, the number of papers in venues, total number venue published in& linear regression, RF, gradient boosted regression tree (GBRT) & $R^{2}$, MAPE& h-index \\ \hline
    ~\cite{ayaz2018predicting}&current h-index, number of co-authors& number of publications, years since publishing first article, average citations per paper &  &number of distinct journals published in, number of articles in top 10 journals in Computer Science & fitting regression equation& $R^{2}$, RMSE, Max\_error& h-index\\ \hline
  \end{tabular}
  \label{tab:4}
\end{table*}

McCarty et al.~\cite{mccarty2013predicting} integrate the variables reflecting an author's collaborative behavior into regression model for predicting author's h-index. In their studies, the number of authors across all h-index articles, average authors of each article, normalized mean betweenness, average number of articles published between co-authors, and average h-index among co-authors are selected as features to train the learning model. Penner et al.~\cite{penner2013predictability} propose an age-dependent cumulative model, and the predictive power of this model depends on scholars' career age.

Dong et al.~\cite{dong2015will} formalize a novel author impact prediction problem to examine the factors driving an article to increase author's h-index. They explore six categories of factors, including author, content of paper, venue, social and temporal factors. According to the correlation analysis of these factors, They find that the author's authority on the paper topic and the venue are important factors to improve the author's h-index. Furthermore, Dong et al.~\cite{dong2016can} find that the impact prediction of a scholar with a higher h-index is more difficult than a scholar with a lower h-index in the future. Ayaz et al.~\cite{ayaz2018predicting} consider a comprehensive data set in the Computer Science from Arnetminer, and explore the effect of different career ages on predicting author's h-index .

The prediction of academic rising stars has attracted widespread attention in academia. Daud et al.~\cite{daud2015using} first try to use supervised machine learning technologies for predicting the rising stars. A set of features are constructed on basis of scholars and their social attributes, such as author influence, author contribution, venue citation and co-author citations. Weihs et al.~\cite{weihs2017learning} generate a collection of 44 features for each author, and integrate these features into several regression models such as linear regression, simple Markov, Random Forest (RF)~\cite{breiman2001random}, and Gradient Boosted Regression Trees (GBRT)~\cite{mohan2011web}.

\subsubsection{Network-based predictive model}
Li et al.~\cite{li2014network} propose a network-base model with two parameters for finding the influential scholars. They use Katz\-Bonacich centrality to define scholarly network prestige~\cite{ballester2006s}. The parameter $\alpha$ shows the useful information is exogenous to the scholarly networks, and parameter $\beta$ measures the robustness of experimental results.
Based on the co-author networks, Daud et al.~\cite{daud2017finding} develop a weighted mutual influence rank (WMIRank) for finding academic rising stars by combining three attributes of co-authorship such as co-author's citations based mutual influence, co-author' order based mutual influence and co-author venues' citations based mutual influence. Zhang et al.~\cite{zhang2016rising} propose ScholarRank method by considering three factors: citation count of authors, the mutual influence among co-authors and the mutual reinforce process of different academic entities in the heterogeneous academic networks. These academic networks include citation networks, paper-journal networks and paper-author networks. However, most network-based predictive models for predicting author impact ignore an important fact that the academic networks evolve over time.

Zhang et al.~\cite{zhang2018pepsi} propose the PePSI method of individualized predictive scholars' influence in the time series academic network. They mainly classify scholars into different types according to their citation dynamics. Furthermore, they construct four academic networks: temporal paper citation networks, temporal co-author networks, temporal paper-venue networks and temporal paper-author networks. Based on these academic networks, they calculate each scholar's impact by applying the random walk algorithms.

\subsubsection{Generative predictive model}
Although the feature-driven and network-based predictive models can improve the accuracy of the author impact prediction to a certain extent, these models lack explanatory power. Sinatra et al.~\cite{Sinatra2016Quantifying} quantify scholar impact by formulating a stochastic model that assigns an unique individual parameter $Q$ to each scholar. The $Q$ value reflects an author's influence on a paper impact, and it is a constant in a scholar's career. The $Q$ parameter for scholar $i$ is define as:
\begin{equation}\label{eq:e8}
  Q_{i}=e^{ \left \langle  \log{c_{i\alpha}} \right \rangle - \mu_{p}}
\end{equation}
where $Q_{i}$ represents the $Q$ value of scholar $i$. $\left \langle  log c_{i\alpha}\right \rangle$ represents the average logarithmic citations of all papers published by scholar$i$. $\alpha$ represents scholar $i$'s $\alpha$-th paper. $\mu_{p}$ is equal to $\left \langle \widehat{p} \right \rangle$. They find that a scholar's h-index is jointly determined by the $Q$ parameter and the productivity $N$. In addition, they find that scholar's future career impact can be predicted by the $Q$ value. The $Q$ model can be explained by temporal changes in productivity, luck, and heavy tailed nature of a scholar's impact distribution.

\subsection{Evaluation indices}
In this subsection, we introduce several evaluation metrics to verify the validity of author impact prediction. Except for Precision and Recall, Mean Absolute Error (MAE), Root Mean Square Error (RMSE), F-Measure and Accuracy are used as evaluation metrics. The MAE can quantify how close between the predictions and the ground truth is. It is defined as $MAE=\frac{1}{n}\sum_{i=1}^{n}\left |{e_{i}}  \right |$, where MAE is an average of the absolute errors $\left |{e_{i}}  \right |$ , which is equal to $\left |f_{i} -y_{i}\right |$, $f_{i}$ is the prediction, $y_{i}$ is the true value, and $n$ represents the number of predictions. RMSE is similar to MAE, which is defined as : $RMSE=\sqrt{\frac{1}{n}\sum_{i=1}^{n}{e_{i}}^{2}}$, RMSE also provides the average error and quantifies the overall error rate. F-Measure is defined as $F-Measure=\frac{(\beta ^{2}+1)PR}{\beta ^{2}P+R}$, where $\beta$ represents parameter. $P$ is the accuracy rate (Precision), and $R$ is the recall rate (Recall). Accuracy shows the fraction of papers correctly predicted for a given error tolerance $\epsilon$. This metric is defined as: $Accuracy=\frac{1}{n} \sum_{i=1}^n \mid\frac{\left |e_{i}  \right |}{y_{i}}\leq\epsilon\mid$.

\section{Open issues and challenges}
In this section, we show several open issues for further research in this area, including author impact inflation, unified evaluation standards, academic success gene, identifying the origins of hot streaks, and higher-order academic networks analysis.

\subsection{Author impact inflation}
The consideration of author impact inflation, which is mainly caused by citation inflation, is important in the measurement, interpretation, and modeling of science. The citation inflation stems from the exponential growth of scholarly papers, and affects the relative number of citations~\cite{pan2018memory}. Further, citation inflation influences the comparative evaluation of scholars, institutions, and countries across different periods. For this reason, normalization strategies for quantifying citation impact between disciplines are consistently explored in the bibliometrics community~\cite{Waltman2016A}. As author impact is related to author's citations, citation inflation has increased the difficulty of author impact evaluation and prediction.


\subsection{Unified evaluation standards}
Although the predictive modeling of author impact has generated enormous progress in quantifying scientific studies, different researchers choose different predictive performance metrics. For example, Ayaz et al.~\cite{ayaz2018predicting} choose $R^{2}$ and RMSE to evaluate the predictive results, whereas Dong et al.~\cite{dong2016can} decide to measure the Precision, Recall, F1-score, AUC, Accuracy, MAP and Pre@3. To more objectively qualify scientific studies, there's a need for defining a unified evaluation standard.

\subsection{Academic success gene}
In the past, more attention has focused on predicting author's h-index and academic rising stars by using feature-driven model and network-based model. Yet, little is known about the mechanisms of the temporal evolution of author impact. Although $Q$ parameter can accurately predict a scholar's impact, the dependence of $Q$ on exogenous factors, such as education level, current institution, or publication habits, remains unknown~\cite{Sinatra2016Quantifying}. More likely, the academic success genes include multiple factors, rather than a single one. Uncovering the origin of the academic success genes is a challenging task, which not only could offer a better understanding of evolution of scholar impact, but also might guide and train high-impact scholars.

\subsection{Identifying the origins of hot streaks}
The hot streak phenomena in scientists' individual careers has attracted researchers' attentions. Liu et al.~\cite{liu2018hot} uncover that hot streaks fundamentally can drive the collective academic impact of a scholar. The uncovered hot streak phenomena are particularly crucial to understanding the long-term academic impact of a scholar in his/her career. If we ignore it, the future impact of a scholar's career will be overestimated or underestimated. They show a hot-streak model that captures a real wide range of academic impact trajectories of a scholar. However, the origins of hot steaks phenomena remain unknown.

\subsection{Higher-order academic networks analysis}
Researchers have traditionally focused on analyzing homogeneous and heterogeneous academic networks to quantify the impact of scholars. Most of the prior studies reflect the citation dynamics by first-order academic networks, including two nodes: the citing nodes and the cited nodes. Due to the first-order academic networks cannot reflect the true citation flow pattern, and the higher-order citation networks can more accurately represent the citation dynamics, the higher-order academic networks analysis can help us understand the long-term impact of scholars in their careers ~\cite{bai2018quantifying}. As the analysis of higher-order academic networks is difficult because its complexity of constructing the higher-order dependencies in academic networks, further study on this topic remains an open challenge in scholar impact evaluation.

\section{Conclusion}
In this paper, we have provided a comprehensive review on author impact evaluation and prediction, focusing on different quantifying methods that can be used for author impact evaluation and prediction. Several changes have taken place in this area: (1) from simple analysis to prediction; (2) from single-dimensional assessment to multi-dimensional assessment; (3) from explicit features to implicit features; (4) from unstructured metrics to structured metrics. However, the analysis of the literature on author impact evaluation and prediction has led to the conclusion that despite a number of methods have been proposed to resolve the problems in this area, the solutions of some important issues remain unknown, such as author impact inflation, unified evaluation standards, academic success gene, identifying the origins of hot streaks, and higher-order academic networks analysis.

\bibliographystyle{IEEEtran}

\EOD

\end{document}